\documentclass{article}

\usepackage{graphicx,float}
\usepackage[latin1]{inputenc}
\usepackage{mathrsfs,xspace}
\usepackage{amsmath}

\advance\textheight by 1cm
\advance\topmargin -0.2cm

\newtheorem{theorem}{Theorem}
\newtheorem{definition}{Definition}
\newtheorem{lemma}{Lemma}
\newtheorem{corollary}{Corollary}

\newcounter{noqed}
\newcommand{\qed}{ \ifmmode\mbox{ }\fi\rule[-.05em]{.3em}{.7em}\setcounter{noqed}{0}}
\newenvironment{proof}[1][{}]{\noindent{\bf Proof#1. }\setcounter{noqed}{1}}{\ifnum\value{noqed}=1\qed\fi\par\medskip}

\advance\textwidth by 2cm
\advance\oddsidemargin by -1cm
\advance\evensidemargin by -1cm

\floatstyle{ruled}
\newfloat{Algorithm}{htpb}{alg}

\newcommand{\exit}[1]{\operatorname{exit}(#1)}

\newcommand{\Pref}[1]{\operatorname{Pref}(#1)}
\newcommand{\pred}{\operatorname{pred}}

\newcommand{\fbs}{\operatorname{FBS}}

\newcommand{\N}{\mathbf N}

\newcommand{\?}{\mskip1.5mu}

\newcounter{prgline}
\newcommand{\pl}{\\\theprgline\addtocounter{prgline}{1}\>}

\newcommand{\IF}{\text{\textbf{if}}\xspace}
\newcommand{\FI}{\text{\textbf{fi}}\xspace}

\newcommand{\DO}{\text{\textbf{do}}\xspace}
\newcommand{\OD}{\text{\textbf{od}}\xspace}

\newcommand{\THEN}{\text{\textbf{then}}\xspace}
\newcommand{\ELSE}{\text{\textbf{else}}\xspace}

\newcommand{\RETURN}{\text{\textbf{return}}\xspace}

\newcommand{\WHILE}{\text{\textbf{while}}\xspace}

\newcommand{\COMMENT}[1]{\ \ \sffamily\textbraceleft\,#1\,\textbraceright}

\newcommand{\lrange}{\operatorname{left}}
\newcommand{\rrange}{\operatorname{right}}
\newcommand{\extent}{\operatorname{extent}}

\def\..{\,\mathpunct{\ldotp\ldotp}} % Middle stuff for intervals. Usage: \..

\renewcommand{\epsilon}{\varepsilon}
\renewcommand{\phi}{\varphi}

\title{Predecessor search with distance-sensitive query time} 
\author{
Djamal Belazzougui\\
Universit\'e Paris Diderot---Paris 7, France\\
Paolo Boldi\quad Sebastiano Vigna\\
Universit\`a degli Studi di Milano, Italy}

\date{}

\begin{document}

\maketitle
\thispagestyle{empty}

\bibliographystyle{splncs03}

\begin{abstract}
A \emph{predecessor (successor) search} finds the largest element $x^-$ smaller
than the input string $x$ (the smallest element $x^+$ larger than or equal to
$x$, respectively) out of a given set $S$; in this paper, we consider
the static case (i.e., $S$ is fixed and does not change over time) and assume that
the $n$ elements of $S$ are available for inspection. We present a number of
algorithms that, with a small additional index (usually of $O(n\log w)$ bits,
where $w$ is the string length), can answer predecessor/successor queries
quickly and with time bounds that depend on different kinds of \emph{distance},
improving significantly several results that appeared in the recent literature.
Intuitively, our first result has a running time that depends on the distance
between $x$ and $x^\pm$: it is especially efficient when the input $x$ is
either very close to or very far from $x^-$ or $x^+$; our second result depends on some global notion of distance in the set $S$,
and is fast when the elements of $S$ are more or less equally spaced in the universe; finally, for our third result
we rely on a \emph{finger} (i.e., an element of $S$) to improve upon the first
one; its running time depends on the distance between the input and the
finger.
\end{abstract}

\section{Introduction}

In this paper we study the \emph{predecessor problem} on a \emph{static} set
$S$ of binary strings of length $w$. It is known from recent
results~\cite{PaTTSTOPS,PaTRDNHSP} that structures \emph{\` a la} van Emde Boas
(e.g., y-fast tries~\cite{WilLLWRQ}) using time $O(\log w)$ are optimal among those
using linear space. The lower bound proved in~\cite{PaTTSTOPS,PaTRDNHSP} has
actually several cases, another one is realised, for instance, in
exponential trees~\cite{AnTDOSEST}. A very comprehensive discussion 
of the literature can be found in Mihai P{\v a}tra{\c s}cu's
thesis~\cite{PatLBTDS}.

Albeit the match between upper and lower bounds settles up the problem in the
worst case, there is a lot of space for improvement in two directions: first of
all, if access to the original set $S$ (as a sorted array) is available, 
it is in principle possible to devise an index using \emph{sublinear} additional space
and still answer predecessor queries in optimal time; second, one might
try to improve upon the lower bound by making access time dependent on the
structure of $S$ or on some property relating the query string $x$ to the set
$S$.

In this paper, we describe sublinear indices that
provide significant improvements over previous bounds\footnote{Our space bounds
are always given in terms of the additional number of bits besides those that
are necessary to store $S$.}. Given a set $S$, we denote with $x^-$ and $x^+$
the predecessor and successor in $S$ of a query string $x$, and let $d(x,S) =
\min\{x^+-x,x-x^-\}$ and $D(x,S) = \max\{x^+-x,x-x^-\}$. Note that $d(x,S)$ is small when $x$ is close to some element of $S$, whereas $w-D(x,S)$
is small when $x$ is far from at least one of $x^\pm$.
Finally, let $\Delta_M$ and $\Delta_m$ be the maximum and minimum distance, respectively,
between two consecutive elements of $S$.

\begin{enumerate}
  \item We match the static worst-case search time $O(\log\log d(x,S))$
  of~\cite{BDDFLSUBU}, which was obtained using space $O(n w \log
  \log w)$, but our index requires just $O(n\log w)$ additional space (and
  thus overall linear space).
  \item We improve exponentially over \emph{interval-biased search
  trees}~\cite{BLRRAGCS}, answering predecessor queries in time\footnote{The
  bound in~\cite{BLRRAGCS} is $O(w-\log(x^+-x^-))$. Our proofs are correct
  even replacing $D(x,S)$ with $x^+-x^-$, but the difference is immaterial as
  $x^+-x^-\leq 2D(x,S)$, and we like the duality with the previous bound
  better.} $O(\log(w-\log D(x,S)))$, again using just $O(n\log w)$ additional bits of
  space.
  \item We improve exponentially over \emph{interpolation
  search}~\cite{DJPISND}, answering
  predecessor queries in time $O(\log\log(\Delta_M/\Delta_m))$,
  always using just $O(n\log w)$ additional bits of space.
  \item Finally, with slightly more (but still sublinear) space we can exploit a
  \emph{finger} $y\in S$ to speed up our second result to $O(\log(\log|x-y|-\log D(x,S)))$, which is in
  some cases better than the bound reported
  in~\cite{AnTDOSEST}, and improves exponentially over
  interval-biased search trees, which need time  $O(\log(2^w-y)-\log
  D(x,s))$~\cite{BLRRAGCS}.
\end{enumerate}

We remark that combining the first two results we show that predecessor search can be 
performed in time  $O(\log \min \{\,\log d(x,S),w-\log D(x,S)\,\})$ using $O(n\log w)$ 
bits of additional space. Our results are obtained starting from a refined version of
\emph{fat binary search in a z-fast trie}~\cite{BBPMMPH} in which the initial search interval
can be specified under suitable conditions, confirming the intuition that fat
binary search can be used as a very versatile building block for data structures.

\section{Notation and tools}
\label{sec:notation}

%In this section, we introduce some terminology and notation adopted throughout
%the rest of the paper. 
We use von Neumann's definition
and notation for natural numbers, and identify $n=\{\?0, 1, \ldots, n-1\?\}$, so
$2=\{\?0,1\?\}$ and $2^*$ is the set of all binary strings.
If $x$ is a string, $x$ juxtaposed with an interval is the substring of $x$ with
those indices (starting from 0). Thus, for instance, $x[a\..b)$
is the substring of $x$ starting at position $a$ (inclusive) and ending at position
$b$ (exclusive). We will write $x[a]$ for $x[a\..a]$. The symbol $\preceq$
denotes prefix order, and $\prec$ is its strict version. Given a prefix $p$, we
denote with $p+1$ and $p-1$ the strings in $2^{|p|}$ that come before and after
$p$ in lexicographical order; in case they do not exist, we assume by convention
that the expressions have value $\bot$. 
All logarithms in this paper are binary and we postulate that $\log x=1$
whenever $x<2$.

Given a set $S$ of $n$ binary strings of length $w$, we let
\begin{align*}
	x^- &= \max\{\?y \in S \mid y< x\?\} &&\text{(the \emph{predecessor} of
	$x$ in $S$)}\\
	x^+ &= \min\{\?y \in S \mid y\geq x\?\} && \text{(the \emph{successor} of
	$x$ in $S$)},
\end{align*}
where $\leq$ is the lexicographic order. A \emph{predecessor/successor} query is
given by a string $x$, and the answer is $x^\pm$. In this paper, for the sake
of simplicity we shall actually concentrate on predecessor search only,
also because our algorithms actually return the \emph{rank} of the predecessor
in $S$, and thus are in principle more informative (e.g., the successor can be
immediately computed adding one to the returned index).

We assume to be able to store a constant-time $r$-bit function on $n$ keys using
$rn+cn +o(n)$ bits for some constant $c\geq 0$: the function may return
arbitrary values outside of its domain (for practical implementations
see~\cite{BBPTPMMPH}).

We work in the standard RAM model with a word of length $w$, allowing
multiplications, and adopt the full randomness assumption. Note, however, 
that the dependence on multiplication and full randomness is only due to the
need to store functions succinctly; for the rest, our algorithms do not depend
on them.

\subsection{Z-fast tries} 

We start by defining some basic notation for compacted tries. Consider the compacted
trie~\cite{KnuACP} associated with a prefix-free set of strings\footnote{Albeit
the results of this paper are discussed for sets of strings
of length $w$, this section provides results for arbitrary sets of prefix-free strings whose length is
$O(w)$.} $S\subseteq 2^*$. \begin{figure}[t]
\centering
\includegraphics[scale=.80]{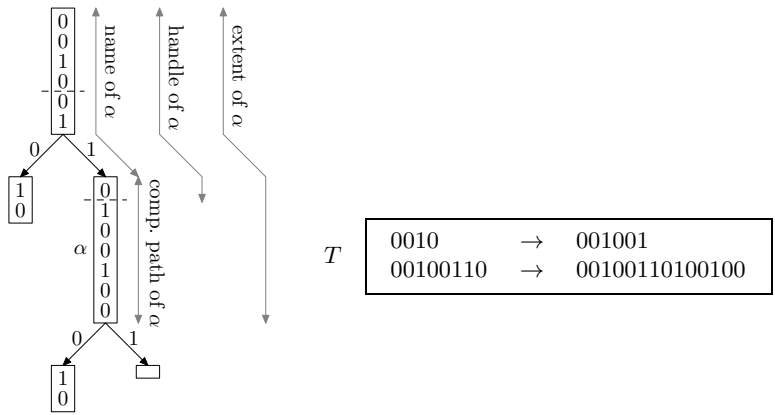}\qquad\raisebox{2cm}{\small
$T$~\begin{tabular}{c}
\fbox{\begin{tabular}{lcl}
0010 & $\to$ & $001001$\\
00100110 & $\to$ & $00100110100100$\\
\end{tabular}}\qquad
\end{tabular}
}
\caption{\label{fig:ztrie}(above) A compacted trie, the related names, and the
function $T$ of the associated z-fast trie. The skip interval for $\alpha$ is
$[7\..13]$. Dashed lines show the end of the handles of internal nodes.}
\end{figure}
Given a node $\alpha$ of the trie (see Figure~\ref{fig:ztrie}):
\begin{itemize}
	\item the \emph{extent} of $\alpha$, denoted by $e_\alpha$, is the longest
	common prefix of the strings represented by the external nodes 
	that are descendants of $\alpha$ (extents of internal nodes are called
	\emph{internal extents});
	\item the \emph{compacted path} of $\alpha$, denoted by $c_\alpha$, is the
	string labelling $\alpha$;
 	\item the \emph{name} of $\alpha$ is the extent of $\alpha$
 	deprived of its suffix $c_\alpha$.
	\item the \emph{skip interval} of $\alpha$ is $[1\..|e_\alpha|]$ for the root, and $[|n_\alpha|\..|e_\alpha|]$
	for all other nodes.
\end{itemize}

Given a string $x$, we let $\exit x$ be the exit node of $x$, that is, the
only node $\alpha$ such that $n_\alpha$ is a prefix of $x$ and
either $e_\alpha=x$ or $e_\alpha$ is not a prefix of $x$.
We recall a key definition from~\cite{BBPMMPH}:

\begin{definition}[2-fattest numbers and handles] 
\label{def:twofattest}
The \emph{2-fattest number} of an interval
$(a\..b)$ of positive integers is the unique integer in $(a\..b)$ that is
divisible by the largest power of two, or equivalently, that has the largest
number of trailing zeroes in its binary representation. The \emph{handle}
$h_\alpha$ of a node $\alpha$ is the prefix of $e_\alpha$ whose length is 2-fattest number in the skip interval of $\alpha$
(see Figure~\ref{fig:ztrie}). If the skip interval is empty (which can only
happen at the root) we define the handle to be the empty string.
\end{definition}
We remark that if $f$ is 2-fattest in $(a\..b)$, it is also 2-fattest in every
subinterval of $(a\..b)$ that still contains $f$.

\begin{definition}[z-fast trie]
Given a prefix-free set $S\subseteq 2^*$, the \emph{z-fast trie on $S$} is 
a function $T$ mapping $h_\alpha \mapsto e_\alpha$, for each internal node
$\alpha$ of the compacted trie associated with $S$, and any other string to an
arbitrary internal extent.
\end{definition}

The most important property of $T$ is that it makes us able to find very quickly
the name of the exit node of a string $x$ using 
a fat binary search (Algorithm~\ref{algo:query}). The basic idea is that of locating the
longest internal extent $e$ that is a proper prefix of $x$:
the name of $\exit x$ is then $x[0\..|e|+1)$. The algorithm
narrows down an initial search interval by splitting it on its 2-fattest number (rather than on its midpoint).
The version reported here (which builds upon~\cite{BBVDZT}) has two main features:
very weak requirements on $T$, and the possibility of starting the search on a small interval. The
latter feature will be the key in obtaining our main results.

\begin{Algorithm}
\smallskip\textbf{Input:} a nonempty string $x\in2^*$, an integer $0\leq
a<|x|$ such that $a=0$ or $x[0\..a)$ is an internal extent of the compacted trie on $S$, and
an integer $b\leq|x|$ larger than the length of the longest internal extent of the
compacted trie on $S$ that is a proper prefix of $x$.

\textbf{Output:} the name of $\exit{x}$	
\vspace{-1.5em}
\begin{tabbing}
\setcounter{prgline}{0}
\hspace{0.5cm} \= \hspace{0.3cm} \= \hspace{0.3cm} \= \hspace{0.3cm} \= \hspace{0.3cm} \= \kill
\pl\WHILE $b-a>1$ \DO
\pl\>$f\leftarrow $ the 2-fattest number in $(a\..b)$
\pl\>$e\leftarrow T(x[0\..f))$
\pl\>\IF $f\leq|e|\land e\prec x$ \THEN $a\leftarrow |e|$ \COMMENT{Move from $(a\..b)$ to $(|e|\..b)$}
\pl\>\ELSE $b\leftarrow f$ \COMMENT{Move from $(a\..b)$ to $(a\..f)$}
\pl\OD
\pl\IF $a=0\land e_\text{root}\neq\epsilon$ \RETURN $\epsilon$
\pl\ELSE \RETURN $x[0\..a+1)$
\end{tabbing}
\caption{\label{algo:query}Fat binary search on the z-fast trie: at the end of the
execution we return the name of $\exit x$.}
\end{Algorithm}

%We will use the following property of 2-fattest numbers, proved
% in~\cite{BBPMMPH}:

\begin{lemma}\label{lem:correctness}
Let $p_0=\epsilon$ and $p_1$,~$p_2$,
$\dots\,$,~$p_t$ be the internal extents of the compacted
trie that are \emph{proper} prefixes of $x$, ordered by increasing length. 
Let $(a\..b)$ be the interval maintained by 
Algorithm~\ref{algo:query}. Before and after each iteration the following invariants 
are satisfied: 
\begin{enumerate}
    \item\label{enu:lema} $a=|p_j|$ for some $j$;
    \item\label{enu:lemb} $|p_t|< b$.
\end{enumerate}
Thus, at the end of the loop, $a=|p_t|$.
\end{lemma}
\begin{proof}
\noindent(\ref{enu:lema})
The fact that $a=|p_j|$ for some $j$ is true at the beginning, and when
$a$ is reassigned (say, $a \leftarrow |e|$) it remains true: indeed, since $e$
is an internal extent, $a<f\leq |e|$ and $e\prec x$, $e=p_k$ for some $k>j$.

\noindent(\ref{enu:lemb})
By (\ref{enu:lema}), $a$ is always the length of some $p_j$,
so $b>|p_t|$ at the beginning, and then it can only decrease; thus,
$(a\..b)$ contains the concatenation of some contiguous skip intervals of the
proper ancestors of $\exit x$ up to the skip interval of $\exit x$ (which may
or may not be partially included itself).

Now, assume by contradiction that when we update $b$ there is a node $\alpha$
with extent $e_\alpha$ which is a proper
prefix of $x$ of length $f$ or greater. Since $f$ is 2-fattest in $(a\..b)$, it
would be 2-fattest in the skip interval of $\alpha$ (as the latter is contained
in $(a\..b)$), so $x[0\..f)$ would be the handle of $\alpha$, and $T$ would have
returned $e_\alpha$, which satisfies $f\leq |e_\alpha|$ and $e_\alpha\prec x$, contradicting the fact that we are
updating $b$. We conclude that the invariant $|p_t|<b$ is preserved.\qed
\end{proof}

\begin{theorem}
\label{thm:correctnessfbs}
Algorithm~\ref{algo:query} completes in at most $\lceil\log(b-a)\rceil$
iterations, returning the name of $\exit x$.
\end{theorem}
\begin{proof}
We first prove the bound on the number of iterations. Note that given an
interval $(\ell\..r)$ in which there is at most one multiple of $2^i$, the two
subintervals $(\ell\..f)$ and $(f\..r)$, where $f$ is the 2-fattest number in
$(\ell\..r)$, contain both at most one multiple of $2^{i-1}$ (if one of the
intervals contained two such multiples, there would be a multiple of $2^i$
inbetween, contradicting our assumption); this observation is \emph{a fortiori}
true if we further shorten the intervals. Thus, we cannot
split on a 2-fattest number more than $i$ times, because at that point the
condition implies that the interval has length at most one. But clearly an interval of length $t$ contains at most one multiple of $2^{\lceil\log t\rceil}$, which shows that the algorithm
iterates no more than $\lceil\log(b-a)\rceil$ times.

Finally, if $t>0$ then $x[0\..|p_t|+1)$ is the name of $\exit x$.
Otherwise, $\exit x$ is the root (hence the special case in Algorithm~1).\qed
\end{proof}

Note that finding the 2-fattest number in an interval requires the computation of
the most significant bit\footnote{More precisely, the 2-fattest number in
$(\ell\..r]$ is $-1 \ll \operatorname{msb}(\ell \oplus r) \mathbin\& r$.}, but
alternatively starting from the interval $(\ell\..r]$
one can set $i=\lceil\log(r-\ell)\rceil$ (this can be computed trivially in time $O(\log(r-\ell))$)
and then check, for decreasing $i$, whether $(-1\ll i)\mathbin\& \ell \neq (-1\ll i)\mathbin\&
r$: when the test is satisfied, there is exactly one multiple of $2^i$ in the interval, 
namely $f=r \mathbin\& -1
\ll i$, which is also 2-fattest. This property is preserved by splitting on 
$f$ and possibly further shortening the resulting interval (see the first part
of the proof of Theorem~\ref{thm:correctnessfbs}), so we can just continue decreasing $i$ and testing, 
which requires still no more than $\lceil\log(r-\ell)\rceil$ iterations.

\subsection{Implementing the function $T$}

A z-fast trie (i.e., the function $T$ defining it) can be implemented in
different ways; in particular, for the purpose of this paper, we 
show that if constant-time access to the elements of $S$ in sorted order is
available, then the function $T$ describing a z-fast trie can be implemented using additional $O(n\log w)$ bits. We will use the
notation $S[i]$ ($0\leq i<|S|$) for the $i$-th element of $S$. We need two key
components:
\begin{enumerate}
  \item a constant-time function $g$ mapping the handles to the length of the
  name of the node they are associated with (i.e., $h_\alpha \mapsto
  |n_\alpha|$ for every internal node $\alpha$);
  \item a \emph{range locator}---a data structure that, given the name of a
  node $n_\alpha$, returns the interval of 
  keys that are prefixed by $n_\alpha$; more precisely, it returns the smallest
  ($\lrange(n_\alpha)$) and largest index ($\rrange(n_\alpha)$, respectively) in $S$ of
  the set of strings prefixed by $n_\alpha$.
\end{enumerate}
The function $g$ can be implemented in constant time using $O(n\log w)$ bits,
and there are constant-time range locators using $O(n\log w)$
bits~\cite{BBPMMPH}.

Now, to compute $T(h)$ for a given handle $h$, we
consider the candidate node name $p = h[0\..g(h))$ and return the longest common
prefix of $S[\lrange(p)]$ and $S[\rrange(p)]$. If $h$ is actually a handle, the whole procedure clearly succeeds and
we obtain the required information; otherwise, we will be returning
some unpredictable internal extent (unless $\lrange(p)=\rrange(p)$, but this
case can be easily fixed). Summing up,
\begin{theorem}
\label{th:zfast}
If access to the set $S$ is available, the z-fast trie can be
implemented in constant time using additional $O(n\log w)$ bits of space.
\end{theorem}
This function enjoys the additional property that, no matter which
the input, it will always return an extent.
We also notice that using the same data it is also easy to implement a function
that returns a node extent given a node name:
\begin{definition}[$\extent$]
Let $p$ be a node name. Then $\extent(p)$ (the extent of the node named $p$)
can be computed in constant time as the longest common prefix of $S[\lrange(p)]$ and $S[\rrange(p)])$.
\end{definition}

\subsection{Using the range locator to check prefixes}

Given a set $P\subseteq \Pref S$, we want to be able to check in constant time
and little space that a prefix $p$ either belongs to $P$, or is not a prefix of a string in $S$. Assume that we have a function $f$ 
defined on $P$ and returning, for each $p\in P$, the length of the name of the exit node of $p$. 
Our key observation is that a range locator, combined with access to the array $S$, can be used to ``patch''
$f$ so that it returns a special value $\bot$ outside of $\Pref S$:
% \footnote{We remark
% that we cannot claim that $f$ will return $\bot$ on elements outside of $P$,
% unless they are not in $\Pref S$ either.} 
\begin{theorem}
\label{th:pref}
Let $P\subseteq\Pref S$ and $f: P\to \N$ be a constant-time function mapping
$p\in P$ to $|n_{\exit p}|$. If access to the set $S$ is available, using an additional
constant-time range locator we can extend $f$ to a constant-time function $\hat f:2^*\to \N
\cup\{\bot\}$ such that $\hat f(p)=|n_{\exit p}|$ for all $p \in P$, and $\hat f(p)=\bot$ for
all $p \not\in\Pref S$.
\end{theorem}
\begin{proof}
To compute $\hat f(p)$ for a $p\in 2^*$ we proceed as follows:
\begin{enumerate}
  \item we compute the candidate length $t=f(p)$ of the name of $\exit p$;
  \item if $t\leq |p|$ and $p\preceq \extent(p[0\..t))$ we return $f(p)$,
  otherwise we return $\bot$.
\end{enumerate}
Clearly, if $p\in P$, by definition $f(p)=|n_{\exit p|}|\leq |p|$, and we
compute correctly the extent of $\exit p$, so we return $f(p)=|n_{\exit p}|$.
On the other hand, if $p \not\in \Pref S$ it cannot be the prefix of an element
of $S$, so in the last step we certainly return $\bot$.
\qed
\end{proof}

% Observe that actually $\hat f$ will return the length of the name of the exit
% node for all prefixes in a set $\hat P$ that satisfies $P\subseteq \hat P\subseteq\Pref S$.

\section{Locally sensitive predecessor search}
\label{sec:pred}

Our purpose is now to combine Theorem~\ref{th:zfast} and~\ref{th:pref} to answer
efficiently predecessor queries in a way that depends on the distance
between the query string and its predecessor and successor. 
First of all, it is clear that we can easily compute the index of the 
predecessor of a string if its exit node is known (e.g., by fat binary search): 
\begin{definition}($\pred$, $\fbs^-$)
Given a string $x$ and the length $t$ of the name of $\exit x$, we define the constant-time function $\pred(x,t)$
as follows:
\begin{itemize}
  \item if $x \preceq \extent(x[0\..t))$, or if the first bit of $x$ at which $x$
  and $\extent(x[0\..t))$ differ is a $0$, 
  the index of the predecessor of $x$ is $\lrange(\exit x)-1$ (we use the
  convention that $-1$ is returned if no predecessor exists);
  \item otherwise, the index of the predecessor of $x$ is $\rrange(\exit x))$.
\end{itemize}
 We denote with $\fbs^-(x,a,b)$ the predecessor index computed by running Algorithm~\ref{algo:query} 
 (with inputs $x$, $a$ and $b$) to obtain the name of $\exit x$ and then invoking $\pred$.
\end{definition}
We remark that the definition above implies that predecessor search
(by means of $\fbs^-(x,0,|x|)$) is possible in time $O(\log w)$ using an index
of $O(n\log w)$ bits.

The rest of this section is devoted at making the computation of the
predecessor of $x$  
more efficient by storing selected prefixes of strings in $S$ 
to reduce significantly the initial search interval of
Algorithm~\ref{algo:query} (i.e., to increase the parameter $a$).
It turns out that this pre-computation phase does dramatically reduce the number
of steps required, making them depend on the distance between the query string
$x$ and its predecessors and successors. More precisely, for a given set $S$ and
a string $x$, let us define
\[
	d(x,S) = \min\{x^+-x,x-x^-\} \quad\text{ and }\quad
	D(x,S) = \max\{x^+-x,x-x^-\};	
\]
if only $x^-$ (equivalently for $x^+$) is defined, we let $d(x,S)=D(x,S)=x-x^-$.
We call $d(x,S)$ (respectively, $D(x,S)$) the \emph{short
distance} (\emph{long distance}) between $x$ and $S$. We will devise two
distinct predecessor algorithms whose performance depend on the short and on the
long distance between the query string and the queried set $S$:
both algorithms use the setup described in Theorem~\ref{th:pref} but with a
different choice of the function $f:P\to \N$.

Before proceeding with the presentation of the algorithms, it is worth observing
the following lemmata:
\begin{lemma}
\label{lemma:hitpref}
	Let $x$ be a string, $j \leq w-\log d(x,S)$ and $p=x[0\..j)$. Then either $p$
	or $p+1$ or $p-1$ belong to $\Pref S$. 
\end{lemma}
\begin{proof}
Suppose that neither $p$ nor $p+1$ nor $p-1$ belong to $\Pref S$; there are
$2^{w-j}$ strings prefixed by $p$ ($x$ being one of them), and the same is true
of $p-1$ and $p+1$. So, the element $y \in S$ that minimises $|y-x|$ (that will be one
of $x^-$ or $x^+$) is such that $|y-x|>2^{w-j}$. Hence $d(x,S)>2^{w-j}$, so
$j>w-\log d(x,S)$, contradicting the hypothesis.\qed
\end{proof}
\begin{lemma}
\label{lemma:shortinprefs}
	Let $x$ be a string; if $p$ is a prefix of $x$ such that $p \in \Pref S$ and
	$|p|>w-\log D(x,S)$, then $x$ is either smaller or larger than all the
	elements of $S$ that have $p$ as prefix.
\end{lemma}
\begin{proof}
Suppose that there is some prefix $p\in \Pref S$ of $x$ longer than $w-\log
D(x,S)$ and that there are two elements of $S$ having $p$ as prefix and that
are smaller and larger than $x$, respectively; in particular, $p$ is also a
prefix of $x^+$ and $x^-$. Since $p$ is the prefix of less than $2^{\log D(x,S)}=D(x,S)$ strings, $x^+-x^-<D(x,S)$; but $x^+-x^-\geq D(x,S)$, so we have a contradiction.\qed
\end{proof}

\subsection{Short-distance predecessor algorithm}

Our first improvement allows for the computation time to depend on short
distances, using techniques inspired by~\cite{BDDFLSUBU}. To this aim, let us
consider the following set of prefixes:
\[
	P=\bigl\{\,x\bigl[0\..w-2^{2^i}\bigr) \mid x \in  S \text{ and }
	i=0,1,\dots,\lfloor\log(\log w - 1)\rfloor\,\bigr\}.
\]
To store the function $f:P\to \N$ needed by Theorem~\ref{th:pref}, we define a subset of $P$:
\[
Q=\bigcup_{\text{node $\alpha$}}\min{}_\preceq\{\,p\in P\mid n_\alpha\preceq p\preceq e_\alpha\,\}
\]
In other words, for every node we take the shortest string in $P$ that sits between the name and the extent
of the node (if any). We can map every element $q\in Q$ to $|n_{\exit q}|$ in space $O(n\log w)$ as $|Q|\leq n$.
Then, we map every $p\in P$ to smallest $i$ such that $p[0\..w-2^{2^i})\in Q$.
This map takes $O(n\log\log w\log\log\log w)=O(n\log w)$ bits. To compute $f(p)$, we first compute the index $i$ using 
the second map, and then query the first map using $p\bigl[0\..w-2^{2^i}\bigr)$. 

Algorithm~\ref{algo:pred-short} probes prefixes of decreasing lengths in the set
$X$. More precisely, at each step we will probe a prefix $p$ of length $t =
w-2^{2^i}$ of the query string $x$; if this probe fails, then $p+1$ and finally
$p-1$ are probed (if they exist). If we succeed in the first case, we have found
a valid prefix of $x$ in the trie, and we can proceed with a fat binary search.
Otherwise, no element is prefixed by $x$, and if by any chance an element is prefixed by $p-1$ or $p+1$ we can easily
locate its predecessor.
 
\begin{Algorithm}
\smallskip\textbf{Input:} a nonempty string $x\in 2^w$

\textbf{Output:} the index $i$ such that $S[i]=x^-$ 
\vspace{-1.5em}
\begin{tabbing}
\setcounter{prgline}{0}
\hspace{0.5cm} \= \hspace{0.3cm} \= \hspace{0.3cm} \= \hspace{0.3cm} \= \hspace{0.3cm} \= \kill
\pl $i \leftarrow 0$
\pl \WHILE $2^{2^i} \leq w/2$ \DO
\pl\> $p \leftarrow x\bigl[0\..w-2^{2^i}\bigr)$
\pl\>$t \leftarrow \hat f(p)$
\pl\>\IF $t \neq \bot$ 
\pl\>\> $e \leftarrow \extent(x[0\..t))$
\pl\>\> \IF $e\prec x$ \RETURN $\fbs^-(x,|e|,|x|)$ \COMMENT{We found a long extent}
\pl\>\> \RETURN $\pred(x,t)$ \COMMENT{We exit at	the node of name $x[0\..t)$}
\pl\> \FI
\pl\> $t \leftarrow \hat f(p+1)$
\pl\> \IF $t \neq \bot$ \RETURN $\lrange((p+1)[0\..t))-1$ \COMMENT{$x^-$ is the
predecessor of $p+1$} \pl\> $t \leftarrow \hat f(p-1)$
\pl\> \IF $t \neq \bot$ \RETURN $\rrange((p-1)[0\..t))$  \COMMENT{$x^-$ is the
successor of $p-1$} \pl\OD
\pl\RETURN $\fbs^-(x,0,|x|)$ \COMMENT{Standard search (we found no
prefix long enough)}
\end{tabbing}
\caption{\label{algo:pred-short}Short-distance speedup.}
\end{Algorithm}

More precisely, it turns out that:
\begin{theorem}
Algorithm~\ref{algo:pred-short} returns the predecessor of $x$
in time $O(\log\log d(x,S))$, and requires an index of $O(n \log w )$
bits of space (in addition to the space needed to store the elements of $S$).
\end{theorem}
\begin{proof}
First we show that the algorithm is correct. If we exit at the first return
instruction, $e$ is a valid extent and a prefix of $x$, so we start correctly a
fat binary search. At the second return instruction we know the $x[0\..t)$ is
the name node $\alpha$, but the extent of $\alpha$ is not a prefix of $x$, so
$x$ exits exactly at $\alpha$, and again we return the correct answer. If $p+1$ is a valid prefix of some element of $S$, but $p$ is not, then the predecessor
of $p$ is the predecessor of the least element prefixed by $p+1$, which we
return (analogously for $p-1$).

By Lemma~\ref{lemma:hitpref}, we will hit a prefix in our set $P$ as soon as
$w-2^{2^i}\leq w-\log d(x,S)$, that is, $i>\log\log\log d(x,S)$. If $i$ is the
smallest integer satisfying the latter condition, then $i-1\leq \log\log\log
d(x,S)$, so $2^{2^i}\leq (\log d(x,S))^2$, which guarantees that the fat binary
search, which starts from an extent of length at least $|e| \geq t \geq
w-2^{2^i} \geq w-(\log d(x,S))^2$, will complete in time $O(\log
b-a)=O(\log\log d(x,S))$ (see Theorem~\ref{thm:correctnessfbs}). If we exit from the loop,
it means that $i>\log\log\log d(x,S)$ implies $2^{2^i}>w/2$, hence
$(\log d(x,S))^2>w/2$, so the last fat binary search (that takes $O(\log w)$
steps to complete) is still within our time bounds.\qed
\end{proof}

\subsection{Long-distance predecessor algorithm}
\label{sec:long}

We now discuss Algorithm~\ref{algo:pred-long}, whose running time depends on long
distances. Let $P$ be the set obtained by ``cutting''
every internal extent $e_\alpha$ to the length of the smallest power of $2$ (if
any) in the skip interval of $\alpha$; more precisely:
\[
	P=\bigcup_\text{$\alpha$ internal}\{\,e_\alpha[0\..2^k) \mid 2^k \in
	[|n_\alpha|\..|e_\alpha|] \text{ and $k$ is the smallest possible}\,\}.
\]
where $\alpha$ ranges over all nodes. Since this time we
have at most one prefix per node, $|P|=O(n)$, so the function $f$ required by
Theorem~\ref{th:pref} can be stored in $O(n\log w)$ bits. 

Algorithm~\ref{algo:pred-long} keeps track of the length $a$ of an
internal extent that is known to be a prefix of $x$. At each step, we try to
find another extent by probing a prefix of $x$ whose length is the smallest power of two larger than
$a$. Because of the way the set $P$ has been built, we can miss the longest
prefix length at most by a factor of two. 

\begin{theorem}	
\label{thm:pred-long}
Algorithm~\ref{algo:pred-long} returns the predecessor of an input string $x$
in time $O(\log(w-\log D(x,S)))$, and requires an index of $O(n \log w)$ bits of space (in addition to the space needed to store the elements of $S$).
\end{theorem}
\begin{proof}
First we show that the algorithm is correct. It can be easily seen that at each
step $a$ is either 0 or the length of an internal extent that is a prefix of
$x$. Moreover, if there is an internal extent of length at least $m$ that is
a prefix of $x$, then $t\neq\bot$, so we if we exit at the first return
instruction, the fat binary search completes correctly. If $t\neq \bot$,
we know that $x[0\..t)$ is the name of a node $\alpha$
(because $(a\..w)$ is a union of consecutive skip intervals, and the smallest power of two in such $(a\..w)$ is
a fortiori the smallest power of two in a skip interval): if  
$x$ is smaller than the smallest leaf
under $\alpha$ (or larger than the largest such leaf), we immediately know the
predecessor and can safely return with a correct value. The return instruction
at the exit of the loop is trivially correct.

Observe that when $m>w-\log D(x,S)$
either the string $x[0\..m)$ will not be in $\Pref S$ (because of
Lemma~\ref{lemma:shortinprefs}) and thus $t=\bot$, or $x$ will be larger (or
smaller) than every element of $S$ prefixed by $x[0\..t)$, which will cause the
loop to be interrupted at one of the last two if instructions. 
Since $m$ gets at
least doubled at each iteration, this condition will take place in at most $\log(w-\log D(x,S))$ iterations; moreover, $m\leq 2a$ (because there is always a power of 2 in the interval $(a\..2a]$), so the fat binary search in the first
return will take no more than $\log(m-a)\leq \log a\leq \log (w-\log D(x,S))$.
If the loop exits naturally, then there is a prefix of $x$ belonging to $\Pref
S$ and longer than $w/2$, hence $w-\log D(x,S)\geq w/2$ and the fat binary
search at the end of the loop will end within the prescribed time bounds.\qed
\end{proof}

\begin{Algorithm}
\smallskip\textbf{Input:} a nonempty string $x\in 2^w$

\textbf{Output:} the index $i$ such that $S[i]=x^-$ 
\vspace{-1.5em}
\begin{tabbing}
\setcounter{prgline}{0}
\hspace{0.5cm} \= \hspace{0.3cm} \= \hspace{0.3cm} \= \hspace{0.3cm} \= \hspace{0.3cm} \= \kill
\pl $a \leftarrow 0$
\pl\WHILE $a<w/2$ \DO
\pl\> $m \leftarrow \text{least power of 2 in $(a\..w)$}$
\pl\> $t \leftarrow \hat f(x[0\..m))$
\pl\> \IF $t = \bot$ \RETURN $\fbs^-(x,a,m)$ \COMMENT{We obtained the longest possible prefix} 
\pl\> $p \leftarrow x[0\..t)$
\pl\> \IF $S[\lrange(p)]\geq x$ \RETURN $\lrange(p)-1$ 
\pl\> \IF $S[\rrange(p)]<x$ \RETURN $\rrange(p)$
\pl\> $a \leftarrow |\extent(p)|$ \COMMENT{This is a valid extent}
\pl\OD 
\pl\RETURN $\fbs^-(x,a,w)$
\end{tabbing}
\caption{\label{algo:pred-long}Long-distance speedup.}
\end{Algorithm}

Finally, we can combine our improvements for short and long distances, obtaining an algorithm that is
efficient when the input $x$ is either very close to or very far from $x^-$ or $x^+$:
\begin{corollary}
It is possible to compute the predecessor of a string $x$ in a set $S$
in time $O(\log \min \{\,\log d(x,S),w-\log D(x,S)\,\})$, using an index that
requires $O(n \log w)$ bits of space (in addition to the space
needed to store the elements of $S$).
\end{corollary}

\section{Globally sensitive predecessor search}

We can apply Theorem~\ref{thm:pred-long} to improve exponentially over the bound
described in~\cite{DJPISND}, which gives an algorithm whose running time
depends on the largest and smallest distance between the elements of $S$. 
More precisely, let
$\Delta_M$ and $\Delta_m$ be, respectively, the largest and smallest distance
between two consecutive elements of $S$.
\begin{corollary}
\label{cor:deltadelta}
Using an index of $O(n\log w)$ bits, it is possible to answer predecessor
queries in time $O(\log\log(\Delta_M/\Delta_m))$.
\end{corollary}
\begin{proof}
See the appendix.
\end{proof}

\section{Finger predecessor search}

We conclude with a generalisation of long-distance search that builds on previous results~\cite{BBPFPSLSA}.
Using $O(n w^{1/c})$ bits (for any $c$) it
is possible to answer \emph{weak prefix search} queries in constant time. A weak
prefix search query takes a prefix $p$ and returns the leftmost and rightmost
index of elements of $S$ that are prefixed by $p$; if no such element exists,
the results are unpredictable (hence the ``weak'' qualifier), but a single access to the set $S$ is sufficient
to rule out this case and always get a correct result. Thus, we will be
able to compute $\lrange(-)$, $\rrange(-)$ and $\extent(-)$ on arbitrary
elements of $\Pref S$ in constant time. As a consequence, also $\pred(x,t)$ can
be extended so to return a correct value for every $t$ such that
$x[0\..t)\in\Pref S$.

The basic idea of Algorithm~\ref{algo:finger} is that of using a \emph{finger}
$y\in S$ to locate quickly an extent $e$ that is a prefix of $x$ with the guarantee that $w-|e|\leq\log|x-y|$. The extent
is then used to accelerate an algorithm essentially identical
Algorithm~\ref{algo:pred-long}, but applied to a reduced universe (the strings
starting with $e$); the running time thus becomes
$O(\log(w-|e|-\log D(x,S)|)=O(\log(\log|x-y|-\log D(x,S)))$.

\begin{Algorithm}
\smallskip\textbf{Input:} a nonempty string $x\in 2^w$ and a $y\in S$ such that $y<x$

\textbf{Output:} the index $i$ such that $S[i]=x^-$ 
\vspace{-1.5em}
\begin{tabbing}
\setcounter{prgline}{0}
\hspace{0.5cm} \= \hspace{0.3cm} \= \hspace{0.3cm} \= \hspace{0.3cm} \= \hspace{0.3cm} \= \kill
\pl $t\leftarrow \max\{\,s\mid y[0\..s)+1\preceq x\,\}$
\pl $e\leftarrow\extent(y[0\..t)+1)$
\pl\IF $y[0\..t)+1\not\preceq e$ \RETURN $\rrange(y[0\..t))$
\COMMENT{$y[0\..t)+1\not\in\Pref S$} \pl\IF $e\not\prec x$ \RETURN $\pred(x,t)$
\COMMENT{$x$ exits between $y[0\..t)+1$ and $e$} \pl $a \leftarrow 0$
\COMMENT{Now $e\prec x$ and $w-|e|\leq\log|x-y|$} \pl\WHILE $a<(w-|e|)/2$ \DO
\pl\> $m \leftarrow \text{least power of 2 in $(a-|e|\..w-|e|)$}$
\pl\> $p \leftarrow x[0\..m+|e|)$
\pl\> \IF $p\not\in\Pref S$ \RETURN $\fbs^-(x,a+|e|,m+|e|)$	 
\pl\> \IF $S[\lrange(p)]\geq x$ \RETURN $\lrange(p)-1$
\pl\> \IF $S[\rrange(p)]<x$ \RETURN $\rrange(p)$
\pl\> $a \leftarrow |\extent(p)|-|e|$ \COMMENT{This is a valid extent}
\pl\OD 
\pl\RETURN $\fbs^-(x,a+|e|,w)$
\end{tabbing}
\caption{\label{algo:finger}Long-distance finger-search speedup.}
\end{Algorithm}

\begin{theorem}
\label{th:finger}
Algorithm~\ref{algo:finger} returns the predecessor of an input string $x$ given
a finger $y\in S$, with $y<x$, in time $O(\log(\log|x-y|-\log D(x,S)))$ using an 
index of $O(n w^{1/c})$ bits 
of space, for any $c$ (in addition to the space needed to store the elements of $S$).
\end{theorem}
\begin{proof}
See the appendix.
\end{proof}

\hyphenation{ Vi-gna Sa-ba-di-ni Kath-ryn Ker-n-i-ghan Krom-mes Lar-ra-bee
  Pat-rick Port-able Post-Script Pren-tice Rich-ard Richt-er Ro-bert Sha-mos
  Spring-er The-o-dore Uz-ga-lis }

\section*{Appendix}

\begin{proof}(of Corollary~\ref{cor:deltadelta})
We use a standard ``universe reduction'' argument, splitting 
the universe $2^w$ by grouping strings sharing the most significant $\lceil \log
n\rceil$ bits. Each subuniverse $U_i$ has size $2^{w-\lceil \log
n\rceil}=O(2^w/n)$, and we let $S_i=S\cap U_i$. Using a constant-time
prefix-sum data structure we keep track of the rank in $S$ of the smallest
element of $S_i$, and we build the indices that are necessary for
Algorithm~\ref{algo:pred-long} for each $S_i$ (seen as a set of strings of
length $w-\lceil \log
n\rceil$). Thus, we can answer a query $x$ in time $O(\log(w-\lceil \log
n\rceil -\log D(x,S_i))$, where $U_i$ is the subuniverse containing $x$. Now
note that $\Delta_M\geq 2^w/n$, and that $\Delta_m\leq x^+-x^-
=(x^+-x)+(x-x^-)\leq 2D(x,S)\leq 2D(x,S_i)$ (unless $x$ the smallest or the
largest element of $S_i$, but this case can be dealt with in constant time). The
bound follows immediately.\qed
\end{proof}

\begin{proof}(of Theorem~\ref{th:finger})
First we show that the algorithm is correct. 
If we exit at the first return instruction, $y[0\..t)+1$ is
not in $\Pref S$, which implies that $x^-$ is prefixed by $y[0\..t)$, and thus
the output is correct. If we exit at the second return instruction, $x$
exits at the same node as $y[0\..t)+1=x[0\..t)$. Otherwise, $e$ is an extent
that is a proper prefix of $x$, and the remaining part of the algorithm
is exactly Algorithm~\ref{algo:pred-long} applied to the set of strings of $S$ 
that are prefixed by $e$, with $e$ removed (the algorithm is slightly
simplified by the fact that we can test membership to $\Pref S$ and compute
extents for every prefix). Correctness is thus immediate.

All operations are constant time, except for the last loop. Note that as soon as
$m+|e|\geq w-\log D(x,S)$ the loop ends or a prefix of $x$ is found (as in the
proof of Algorithm~\ref{algo:pred-long}), and this requires no more than $\log(w-|e|-\log
D(x,S))$ iterations; moreover, $m\leq 2a$ (because there is always
a power of 2 in the interval $(a\..2a]$), so the fat binary search in the first
return will take no more than $\log(m-a)\leq \log a\leq \log (w-|e|-\log D(x,S))$.
If the loop exits naturally, then there is a prefix $e'$ of $x$ belonging
to $\Pref S$ and longer than $(w+|e|)/2$, hence by
Lemma~\ref{lemma:shortinprefs}, $w-\log D(x,S)\geq (w+|e|)/2$; the fat binary search at the
end takes time $O(\log (w-(a+|e|)))=O(\log (w-|e'|))=O(\log( w/2 -
|e|/2)))=O(\log(w-|e|-\log D(x,S)))$, within the prescribed time bounds.\qed
\end{proof}


\begin{thebibliography}{10}
\providecommand{\url}[1]{\texttt{#1}}
\providecommand{\urlprefix}{URL }

\bibitem{AnTDOSEST}
Andersson, A., Thorup, M.: Dynamic ordered sets with exponential search trees.
  J. Assoc.\ Comput.\ Mach.  54(3), 1--40 (2007)

\bibitem{BBPMMPH}
Belazzougui, D., Boldi, P., Pagh, R., Vigna, S.: Monotone minimal perfect
  hashing: {S}earching a sorted table with {$O(1)$} accesses. In: SODA '09.
  pp. 785--794. ACM Press (2009)

\bibitem{BBPTPMMPH}
Belazzougui, D., Boldi, P., Pagh, R., Vigna, S.: Theory and practise of
  monotone minimal perfect hashing. In: ALENEX 2009. pp. 132--144. SIAM (2009)

\bibitem{BBPFPSLSA}
Belazzougui, D., Boldi, P., Pagh, R., Vigna, S.: Fast prefix search in little
  space, with applications. In: de~Berg, M., Meyer, U. (eds.) Algorithms -
  {ESA} 2010. Lecture Notes in Computer Science, vol. 6346,
  pp. 427--438. Springer (2010)

\bibitem{BBVDZT}
Belazzougui, D., Boldi, P., Vigna, S.: Dynamic z-fast tries. In: Ch{\'a}vez,
  E., Lonardi, S. (eds.) SPIRE 2010. Lecture Notes in Computer Science, vol. 6393, pp.
  159--172. Springer (2010)

\bibitem{BLRRAGCS}
Bille, P., Landau, G.M., Raman, R., Sadakane, K., Satti, S.R., Weimann, O.:
  Random access to grammar compressed strings. In: SODA~'11 (2011)

\bibitem{BDDFLSUBU}
Bose, P., Dou\"{\i}eb, K., Dujmovic, V., Howat, J., Morin, P.: Fast local
  searches and updates in bounded universes. In: CCCG2010. pp. 261--264 (2010)

\bibitem{DJPISND}
Demaine, E.D., Jones, T.R., P{\v a}tra{\c s}cu, M.: Interpolation search for
  non-independent data. In: Munro, J.I. (ed.) SODA~'04. pp. 529--530 (2004)

\bibitem{KnuACP}
Knuth, D.E.: The Art of Computer Programming. Addison--Wesley (1973)

\bibitem{PatLBTDS}
P{\v a}tra{\c s}cu, M.: Lower bound techniques for data structures. Ph.D.
  thesis, Massachusetts Institute of Technology, Dept. of Electrical
  Engineering and Computer Science (2008)

\bibitem{PaTTSTOPS}
P{\~a}tra{\c{s}}cu, M., Thorup, M.: Time-space trade-offs for predecessor
  search. In: STOC~'06. pp. 232--240. ACM Press (2006)

\bibitem{PaTRDNHSP}
P\v{a}tra\c{s}cu, M., Thorup, M.: Randomization does not help searching
  predecessors. In: SODA '07. pp. 555--564. SIAM, Philadelphia, PA, USA (2007)

\bibitem{WilLLWRQ}
Willard, D.E.: Log-logarithmic worst-case range queries are possible in space
  {$\Theta(N)$}. Inform.\ Process.\ Lett.  17(2),  81--84 (1983)

\end{thebibliography}
\end{document}